\documentclass[aps,preprint]{revtex4}%
\usepackage{amsfonts}
\usepackage{amsmath}
\usepackage{amssymb}
\usepackage{graphicx}%
\providecommand{\U}[1]{\protect\rule{.1in}{.1in}}

\begin{document}
\title[ ]{Relativity and Radiation Balance for the Classical Hydrogen Atom in Classical
Electromagnetic Zero-Point Radiation}
\author{Timothy H. Boyer}
\affiliation{Department of Physics, City College of the City University of New York, New
York, New York 10031}

\begin{abstract}
When teaching modern physics, every instructor must deal with the apparent
failure of classical physics to prevent the radiation collapse of the nuclear
atom. \ Here we review the understanding of the classical hydrogen atom in
classical electromagnetic zero-point radiation, and emphasize the importance
of special relativity. \ The crucial missing ingredient in earlier
calculational attempts (both numerical and analytic) is the use of valid
approximations to the full relativistic analysis. \ It is pointed out that the
nonrelativistic time Fourier expansion coefficients given by Landau and
Lifshitz are in error as the electromagnetic description of a charged particle
in a Coulomb potential, and, because of this error, Marshall and Claverie's
conclusion regarding the failure of radiation balance is invalid. \ Rather,
using Marshall and Claverie's calculations, but restricted to lowest
nonvanishing order in the orbital eccentricity (where the nonrelativistic
orbit is a valid approximation to the fully relativistic electromagnetic
orbit) radiation balance for classical electromagnetic zero-point radiation is
shown to hold at the fundamental frequencies and associated first overtones.

\end{abstract}
\maketitle

\section{Introduction}

\subsection{\textquotedblleft Modern Physics\textquotedblright\ and the
Apparent Failure of \textquotedblleft Classical Physics\textquotedblright}

In introductory classes in modern physics, an instructor often reviews the
alleged failures of classical physics to account for the blackbody radiation
spectrum and the stability of the hydrogen atom, among other phenomena. \ The
subject is often taught as though the conclusions were unambiguous and closed;
quantum theory is the only alternative. \ However, the matter may not be so
completely closed as we had been led to think. \ Indeed, some of the claims
made in the modern physics texts\cite{mod} regarding the failures of classical
physics are actually wrong. \ In a previous historical review, the blackbody
radiation problem was treated,\cite{bb} and it was pointed out that the
historical arguments are modified in the presence of zero-point radiation in
such a way as to give natural classical explanations for the Planck spectrum.
\ In this article, we return to the problem of the collapse of the nuclear
atom in classical electromagnetism which appeared to be settled in the first
quarter of the 20th century. \ As suggested here, the problem seems
complicated, and instructors actually need to be cautious when describing the
classical physics.

\subsection{History of the Classical Hydrogen Atom Problem}

Contrary to the statements in the modern physics texts,\cite{mod} it has been
suggested for over half a century that it may be possible to understand the
apparently radiationless ground state of the classical hydrogen atom as
arising from a Brownian motion of the electron in classical electromagnetic
zero-point radiation.\cite{M63}\cite{B1970} \ The classical theory of which
this idea forms a crucial part is often referred to as \textquotedblleft
stochastic electrodynamics.\textquotedblright\ During the 1960s, this hopeful
understanding was encouraged by easy calculations involving either perfect
conductors or harmonic oscillators in classical zero-point radiation, which
led to classical derivations in agreement with nature for Casimir forces, van
der Waals forces, specific heats of solids, diamagnetic behavior, and the
Planck blackbody spectrum.\cite{B1975}\cite{DC} \ However, in the 1970s, this
hope was discouraged by calculations which indicated that nonlinear systems
did not preserve radiation balance for zero-point radiation, but rather pushed
the zero-point spectrum toward the Rayleigh-Jeans spectrum.\cite{nonlinear}%
\cite{aav} \ Finally in 1980, the hope of a classical understanding of
hydrogen in zero-point radiation seemed futile because of sophisticated
analytic work by Marshall and Claverie.\cite{MC} \ These authors apparently
showed that a nonrelativistic charged particle in a Coulomb potential in
zero-point radiation did not preserve radiation balance for zero-point
radiation. \ Further analysis indicated that the electron picked up too much
energy from the zero-point radiation, and so the classical hydrogen atom was
self-ionizing.\cite{self} \ Thus the atomic collapse problem of the early
twentieth century was replaced by a new problem where zero-point radiation
provided too much energy. \ Marshall and Claverie regarded their calculations
as being consistent with the earlier nonlinear calculations of the 1970s
leading to the Rayleigh-Jeans spectrum for radiation balance, but depreciated
the previous suggestions that the use of relativistic scatterers might be the
missing ingredient leading back to the hoped-for stability and radiation
balance for hydrogen in zero-point radiation. \ 

The gloom surrounding stochastic electrodynamics was suddenly lifted in 2003
by nonrelativistic numerical simulations carried out by Cole and Zou for the
classical hydrogen atom in classical zero-point radiation.\cite{CZ} \ Using no
adjustable parameters, they showed that the probability distribution for the
position of the nonrelativistic electron seemed to be approaching that given
by the Schroedinger equation ground state. \ However, in 2015, a damper on the
enthusiasm arrived with new and longer calculations by Nieuwenhuizen and Liska
which indicated disagreement with the Schroedinger ground state and which
emphasized the self-ionizing aspects of the simulation calculations.\cite{NL}
\ Although it had been noted that relativity would modify the plunging orbits
where the self-ionization occurred,\cite{B2016} Nieuwenhuizen and Liska
dismissed the role of relativity; they carried out further calculations
suggesting that relativity made only insignificant changes in their
simulations.\cite{NL2} \ 

\subsection{Evidence for the Importance of Relativity}

Recently the outlook has changed yet again due to the first-ever scattering
calculation involving a system which can be regarded as
relativistic.\cite{B2018} \ The first important observation for the new
scattering calculation is that we can think of a nonrelativistic calculation
as a valid approximation to the accurate relativistic calculation in the
parameter range where both calculations agree. \ Specifically, a particle in a
harmonic potential taken in the small-amplitude limit, can be regarded as
approximately relativistic, since when the velocity is small, $v<<c,$ the
particle momentum change $d\mathbf{p}/dt$ and the particle kinetic energy
calculated nonrelativistically agree with the accurate relativistic
calculations, with the first corrections starting at order $\left(
v/c\right)  ^{2}$. \ The second needed aspect for a relativistic scattering
calculation was provided by Huang and Batelaan's emphasis that when the
oscillation amplitude of a harmonic oscillator is non-zero, then the
oscillator will interact with radiation not only at the fundamental frequency
but at all its harmonics.\cite{HB} \ A purely harmonic oscillator of finite
amplitude of oscillation will produce not only dipole radiation but also
quadrupole radiation and indeed radiation at all the harmonics. \ Thus the
purely harmonic oscillator of finite amplitude but with $v<<c$ provides a
valid approximation to a relativistic system but also connects the system to
radiation at several frequencies. \ Here finally was a simple mechanical
system where radiation equilibrium is not forced by the mechanical system but
rather is determined by the multipole electromagnetic interactions connected
to the full space and time dependence of the radiation field. \ And indeed,
calculations showed that for this \textit{relativistic} system, radiation
equilibrium required a \textit{relativistic} spectrum of random
electromagnetic radiation, and \textit{not} the Rayleigh-Jeans
spectrum.\cite{B2018} \ Electromagnetism is a relativistic theory and
Lorentz-invariant classical zero-point radiation should fit with
electromagnetic theory. \ For this relativistic oscillator system, radiation
balance held for a Lorentz-invariant spectrum of random classical radiation,
indeed for classical electromagnetic zero-point radiation. \ In this case,
relativity was indeed the missing ingredient for radiation balance. \ 

\subsection{Reconsideration of the Classical Hydrogen Ground State}

This result involving a relativistic system and classical zero-point radiation
has changed the outlook for classical mechanical systems in random radiation.
\ In the present article, we will review briefly the ideas of classical
zero-point radiation, and then we will look again at a charged particle in a
Coulomb potential in classical zero-point radiation. \ Naively, we tend to
think that nonrelativistic approximations should always be valid for a charged
particle in a Coulomb potential. \ However, we will point out that some of the
earlier work on hydrogen does not provide nonrelativistic limits which are
consistent with relativistic behavior. \ \ We will reconsider the analytic
calculations of Marshall and Claverie which were carried out in 1980. \ Here
we will take as valid precisely those aspects which are consistent with the
relativistic limit, and only those aspects. \ In contradiction with the
earlier claims that radiation balance did not hold, we will show that Marshall
and Claverie's calculations give radiation balance at both the fundamental
orbital frequencies and the associated first overtones for calculations
involving terms through first order in the orbital eccentricity where the
nonrelativistic calculations provide a valid approximation to the actual
relativistic electromagnetic situation. \ Also, we point out that
nonrelativistic numerical simulations may underestimate the importance of
relativistic corrections, especially in regions of large eccentricity. \ 

\section{Basic Ideas of Classical Electrodynamics with Classical
Electromagnetic Zero-Point Radiation}

\subsection{Elements Missing from Classical Theory in 1900}

At the turn of the 20th century, physicists did not appreciate two crucial
aspects of classical theory.\cite{bb} \ 1) Classical electrodynamics is a
relativistic theory and any valid analysis must be consistent with relativity.
\ 2) Lorentz-invariant classical electromagnetic zero-point radiation is
present in nature. \ 

Indeed, there is a profound conflict within classical physics which has not
been appreciated. \ Classical electrodynamics always connects charges to
radiation so that an accelerating charge emits electromagnetic radiation. \ On
the other hand, this connection of a particle with its field is absent in
classical mechanics where a (frictionless) particle in a potential may
oscillate indefinitely without losing energy. \ Because physicists were so
used to the ideas of classical mechanics in the early 20th century, they
attempted to construct a theory which broke the connection between a charged
particle and its radiation field. \ Thus in his description of old quantum
theory, Born writes, \textquotedblleft The endeavour to retain the classical
mechanics as far as possible having proved to be a fertile method, we take as
our first requirement that the stationary states of an atomic system shall be
calculated as far as possible, in accordance with the laws of classical
mechanics, but the classical theory of radiation is
disregarded.\textquotedblright\cite{Born52}

\subsection{Classical Zero-Point Radiation}

The disregard for the classical theory of radiation has been countered and
transformed by the idea of classical electromagnetic zero-point radiation.
\ The first careful and extensive effort to treat problems of atomic physics
entirely within classical electromagnetism was made by Marshall\cite{M63}
beginning in 1963. \ Marshall looked for the classical electromagnetic
radiation field which would give an oscillator the same average energy as
appeared for a quantum oscillator in its ground state. \ This corresponded to
assuming that nature contained random classical radiation at the zero of
temperature with an average energy per normal mode given by $U_{rad\text{zp}%
}(\omega)=(1/2)\hbar\omega$. \ 

Random classical radiation can be written in the form used at the end of the
19th century as a sum over plane waves with random phases in a large cubic box
with sides of length $a,$%
\begin{equation}
\mathbf{E}(\mathbf{r},t)=%
{\displaystyle\sum_{\mathbf{k,\lambda}}}
\widehat{\epsilon}(\mathbf{k},\lambda)\left(  \frac{8\pi U_{rad}(\omega
)}{a^{3}}\right)  ^{1/2}\frac{1}{2}\left\{  \exp\left[  i\mathbf{k}%
\cdot\mathbf{r}-i\omega t+i\theta\left(  \mathbf{k},\lambda\right)  \right]
+cc\right\}  \label{Eran}%
\end{equation}%
\begin{equation}
\mathbf{B}(\mathbf{r},t)=%
{\displaystyle\sum_{\mathbf{k,\lambda}}}
\widehat{\mathbf{k}}\times\widehat{\epsilon}(\mathbf{k},\lambda)\left(
\frac{8\pi U_{rad}(\omega)}{a^{3}}\right)  ^{1/2}\frac{1}{2}\left\{
\exp\left[  i\mathbf{k}\cdot\mathbf{r}-i\omega t+i\theta\left(  \mathbf{k}%
,\lambda\right)  \right]  +cc\right\}
\end{equation}
where the sum over the wave vectors $\mathbf{k}=\widehat{x}2\pi
l/a+\widehat{y}2\pi m/a+\widehat{z}2\pi n/a$ involves integers $l,m,n=0,\pm
1,\pm2,...$ running over all positive and negative values, there are two
polarizations $\widehat{\epsilon}(\mathbf{k},\lambda),$ $\lambda=1,2,$ and the
random phases $\theta(\mathbf{k,\lambda})$ are distributed uniformly over the
interval $(0,2\pi],$ independently for each wave vector $\mathbf{k}$ and
polarization $\lambda$. \ The notation \textquotedblleft$cc$\textquotedblright%
\ refers to the complex conjugate. \ The energy per normal mode at radiation
frequency $\omega=ck$ is given by $U_{rad}(\omega),$ and we have assumed that
the radiation spectrum is isotropic. \ 

\subsection{Time Correlation Function for the Electric Fields}

The correlation functions of the electric and magnetic fields can be obtained
by averaging over the random phases $\theta\left(  \mathbf{k},\lambda\right)
.$ \ Thus for the time-correlation function for the electric field at a single
point in space $\mathbf{r}=0$, we find%
\begin{align}
&  \left\langle E_{i}\left(  0,t\right)  E_{j}\left(  0,t+\tau\right)
\right\rangle \nonumber\\
&  =%
{\displaystyle\sum_{\mathbf{k,\lambda}}}
{\displaystyle\sum_{\mathbf{k}^{\prime}\mathbf{,\lambda}^{\prime}}}
\epsilon_{i}(\mathbf{k},\lambda)\epsilon_{j}\left(  \mathbf{k}^{\prime
},\lambda^{\prime}\right)  \left(  \frac{8\pi U_{rad}(\omega)}{a^{3}}\right)
^{1/2}\left(  \frac{8\pi U_{rad}(\omega^{\prime})}{a^{3}}\right)  ^{1/2}%
\frac{1}{4}\nonumber\\
&  \times\left\langle \left\{  \exp\left[  -i\omega t+i\theta\left(
\mathbf{k},\lambda\right)  \right]  +cc\right\}  \left\{  \exp\left[
-i\omega^{\prime}\left(  t+\tau\right)  +i\theta\left(  \mathbf{k}^{\prime
},\lambda^{\prime}\right)  \right]  +cc\right\}  \right\rangle \nonumber\\
&  =%
{\displaystyle\sum_{\mathbf{k,\lambda}}}
\epsilon_{i}(\mathbf{k},\lambda)\epsilon_{j}\left(  \mathbf{k},\lambda\right)
\left(  \frac{8\pi U_{rad}(\omega)}{a^{3}}\right)  \frac{1}{4}\left\{
\exp\left[  i\omega\tau\right]  +cc\right\}  \label{EiEj}%
\end{align}
since averaging over the random phases involves%
\begin{equation}
\left\langle \exp\left[  i\theta\left(  \mathbf{k},\lambda\right)  \right]
\exp\left[  -i\theta\left(  \mathbf{k}^{\prime},\lambda^{\prime}\right)
\right]  \right\rangle =\delta_{\mathbf{kk}^{\prime}}\delta_{\lambda
\lambda^{\prime}}.
\end{equation}
If the box for the periodic boundary conditions is taken as very large, the
then sums can be replaced by integrals, $%
{\textstyle\sum\nolimits_{\mathbf{k}}}
\rightarrow%
{\textstyle\int}
\left[  a/(2\pi)\right]  ^{3}d^{3}k.$ \ Also, since the radiation is assumed
isotropic, the sum over polarizations $%
{\textstyle\sum\nolimits_{\lambda}}
\epsilon_{i}(\mathbf{k},\lambda)\epsilon_{j}\left(  \mathbf{k},\lambda\right)
=\delta_{ij}-k_{i}k_{j}/k^{2}$ can be replaced by $\delta_{ij}2/3.$ \ Thus the
electric field correlation function (\ref{EiEj}), with $B_{E}\left(
\tau\right)  $ defined as $\left\langle E_{i}\left(  0,t\right)  E_{j}\left(
0,t+\tau\right)  \right\rangle =\delta_{ij}B_{E}\left(  \tau\right)  $,
becomes
\begin{align}
\left\langle E_{i}\left(  0,t\right)  E_{j}\left(  0,t+\tau\right)
\right\rangle  &  =\delta_{ij}%
{\textstyle\int}
\frac{d^{3}k}{\pi^{2}}\frac{2}{3}U_{rad}(\omega)\frac{1}{4}\left\{
\exp\left[  i\omega\tau\right]  +cc\right\} \nonumber\\
&  =\delta_{ij}%
{\textstyle\int\nolimits_{-\infty}^{\infty}}
d\omega\frac{2}{3\pi c^{3}}\omega^{2}U_{rad}\left(  \omega\right)  \exp\left[
i\omega\tau\right]
\end{align}
where we have integrated over angles to give a factor of $4\pi$ and have
extended the $\omega$ integral to $-\infty$ by using the complex conjugate
term. \ We define the spectrum $S_{E}(\omega)$ as%
\begin{align}
S_{E}(\omega)  &  =%
{\textstyle\int\nolimits_{-\infty}^{\infty}}
d\tau B_{E}\left(  \tau\right)  \exp\left[  -i\omega\tau\right] \nonumber\\
&  =%
{\textstyle\int\nolimits_{-\infty}^{\infty}}
d\tau\exp\left[  -i\omega\tau\right]
{\textstyle\int\nolimits_{-\infty}^{\infty}}
d\omega^{\prime}\frac{2}{3\pi c^{3}}\omega^{\prime2}U_{rad}\left(
\omega^{\prime}\right)  \exp\left[  i\omega^{\prime}\tau\right] \nonumber\\
&  =\frac{2}{3c^{3}}2\omega^{2}U_{rad}\left(  \omega\right)  . \label{SEE}%
\end{align}
If we consider zero-point radiation where $U_{rad\text{zp}}\left(
\omega\right)  =\left(  1/2\right)  \hbar\omega,$ then the spectrum is
\begin{equation}
S_{E\text{zp}}\left(  \omega\right)  =\frac{2}{3c^{3}}\hbar\omega^{3}.
\label{SEEzp}%
\end{equation}

\subsection{Dipole Oscillator in Random Classical Radiation \ }

The behavior of a harmonic oscillator system (of natural frequency $\omega
_{0}$, mass $m$, and charge $e$) in random radiation, treated in the dipole
approximation, follows from Newton's second law as\cite{M63}\cite{B1975}%
\begin{equation}
m\ddot{x}=-m\omega_{0}^{2}x-m\tau\dddot{x}\mathbf{+}eE_{x}(0,t) \label{New2nd}%
\end{equation}
where the term in $\dddot{x}$ involves the radiation reaction force with
$\tau=(2e^{2})/(3mc^{3})$. \ The steady-state solution takes the form%
\begin{equation}
x\left(  t\right)  =\frac{e}{m}%
{\displaystyle\sum_{\mathbf{k,\lambda}}}
\epsilon_{x}(\mathbf{k},\lambda)\left(  \frac{8\pi U_{rad}(\omega)}{a^{3}%
}\right)  ^{1/2}\frac{1}{2}\left\{  \frac{\exp\left\{  i\left[  \mathbf{k}%
\cdot\mathbf{r}-\omega t+\theta\left(  \mathbf{k},\lambda\right)  \right]
\right\}  }{-\omega^{2}+\omega_{0}^{2}+i\tau\omega^{3}}+cc\right\}  .
\label{linxt}%
\end{equation}
The successful calculations, involving Casimir forces, van der Waals forces,
specific heats of solids, diamagnetism, and the blackbody
radiation,\cite{B1975}\cite{DC} all involved either purely electromagnetic
fields or else harmonic oscillator systems.

\subsection{Radiation Balance}

When treated in the electric dipole approximation, charged mechanical systems
have a standard interaction with radiation. \ Energy absorption occurs through
the forcing term $e\mathbf{E}\left(  0,t\right)  \cdot\mathbf{v}$ while energy
emission occurs according to Larmor's formula for electric dipole oscillation.
\ \textit{Linear} oscillator systems, treated in the electric dipole
approximation, come to equilibrium when the average energy of the mechanical
oscillator equals the average energy of the normal modes of the radiation
field at the natural frequency $\omega_{0}$ of the linear oscillator.
\ Furthermore, in the dipole approximation, linear oscillator systems will
scatter radiation, but do not change the frequency spectrum of the radiation.
\ Thus since classical electromagnetic zero-point radiation is assumed
isotropic, the harmonic oscillators following Eqs. (\ref{New2nd}) and
(\ref{linxt}) leave the radiation pattern of classical zero-point radiation
undisturbed.\cite{appen} \ Linear oscillators treated in the dipole
approximation preserve radiation balance in any isotropic spectrum of random
radiation, and, in equilibrium, an experimenter would not find any net
radiation emission or absorption at any frequency.

However, as pointed out in Jackson's text,\cite{Jackson} \textquotedblleft
Appreciable radiation in multiples of the fundamental (frequency) can occur
because of relativistic effects ... or it can occur if the components of
velocity are not sinusoidal, even though periodic.\textquotedblright%
\ \ \textit{Nonlinear} oscillators have time Fourier components of velocity
(and hence of radiation) at all multiples of the fundamental oscillation
frequency. \ Thus nonlinear oscillators, treated in the electric dipole
approximation, can absorb energy at one frequency and reradiate the energy at
a different frequency, and so can indeed change the frequency spectrum of the
ambient radiation. \ It was found that a nonrelativistic nonlinear oscillator
did not preserve radiation balance for classical zero-point radiation, but
rather tended to push the spectrum toward the Rayleigh-Jeans spectrum where
the same energy per normal mode exists at all frequencies.\cite{nonlinear}%
\cite{aav} \ Thus, in classical zero-point radiation, the nonlinear
oscillators absorbed radiation at high frequencies where the energy per normal
mode was larger, and reradiated this energy at low frequencies where the
energy per normal mode was smaller, so as to tend to equalize the energy per
normal mode at all frequencies. \ Indeed, within the narrow-line-width
approximation, such scattering calculations for nonlinear systems had been
carried out (but only partly published) by van Vleck in 1924.\cite{VV} \ 

\section{Refinements in the Mathematical Analysis}

\subsection{Use of Action-Angle Variables in Mechanical Systems}

The interest in nonlinear scattering systems brought attention back to the use
of action-angle variables in classical mechanics associated with old quantum
theory.\cite{aav}\cite{Born52} The action variables are adiabatic invariants,
and so are ideal for use in perturbation calculations where the mechanical
system is weakly perturbed by electromagnetic radiation. \ The calculations by
van Vleck\cite{VV} in 1924 and by the physicists interested in zero-point
radiation in the 1970s involved a purely \textit{dipole} connection between
the mechanical system and radiation. \ Thus the purely mechanical system of a
particle of mass $m$ in a potential $V(\mathbf{r})$ was assumed described by
the mechanical equation
\begin{equation}
\frac{d\mathbf{p}}{dt}\mathbf{=-}\nabla V(\mathbf{r}).
\end{equation}
Then action-angle variables $w_{i},J_{i},$ $i=1,2,3$ were introduced so that
the particle position could be described as a multiply periodic system with
expansion coefficients $\mathbf{D}_{lmn}(J_{1},J_{2},J_{3})$ such that
\begin{equation}
\mathbf{r}=%
{\textstyle\sum\nolimits_{lmn}}
\mathbf{D}_{lmn}(J_{1},J_{2},J_{3})\exp\left[  i\left(  lw_{1}+mw_{2}%
+nw_{3}\right)  \right]  \label{rexpan}%
\end{equation}
where the integers $l,m,n=0,\pm1,\pm2,...$, and where the angle variables
advance uniformly in time%
\begin{equation}
w_{i}=\omega_{i}t+\beta_{i}.
\end{equation}
The (angular) frequencies $\omega_{i}$ are functions of the action variables,
$\omega_{i}=\omega_{i}(J_{1},J_{2},J_{3})$, and the action variables
$J_{i},~i=1,2,3$ are constant in time. \ 

\subsection{Mechanical System Perturbed by Random Radiation}

If the particle in the potential $V\left(  \mathbf{r}\right)  $ carries a
charge $e$ in the presence of random radiation, then the equation of motion in
the dipole approximation becomes a generalized Langevin equation%
\begin{equation}
\frac{d\mathbf{p}}{dt}\mathbf{=-}\nabla V(\mathbf{r})+\frac{2e^{2}}{3c^{3}%
}\mathbf{\dddot{r}+}e\mathbf{E}(0,t), \label{Lang}%
\end{equation}
where the term $\left[  2e^{2}/(3c^{3})\right]  \mathbf{\dddot{r}}$ represents
the force due to dipole radiation damping and the term $e\mathbf{E}(0,t)$
represents the force due to the random electric field, again taken in the
dipole approximation $e\mathbf{E}(\mathbf{r},t)\approx e\mathbf{E}(0,t)$. \ 

\subsection{Quasi-Markov Process and the Fokker-Planck Equation}

The charged particle whose motion is described by Eq. (\ref{Lang}) will both
lose energy by radiation emission and absorb energy from the ambient
radiation. \ If the magnitude $e$ of the charge is small, then the particle
will carry out many mechanical oscillations before there is significant energy
exchange with the radiation field. \ In this \textquotedblleft
narrow-line-width\textquotedblright\ approximation, Marshall\cite{M-Br} and
Marshall and Claverie\cite{MC-Br} showed that the charged particle can be
regarded as carrying out a quasi-Markov process described by a Fokker-Planck
equation for the probability distribution $P(J_{1},J_{2},J_{3},t)$ of the
action variables,
\begin{equation}
\frac{\partial}{\partial t}P(J_{1},J_{2},J_{3},t)=%
{\textstyle\sum\nolimits_{i}}
\frac{\partial}{\partial J_{i}}\left(  -A_{i}P(J_{1},J_{2},J_{3},t)+%
{\textstyle\sum\nolimits_{j}}
B_{ij}\frac{\partial}{\partial J_{j}}P(J_{1},J_{2},J_{3},t)\right)  .
\label{FP}%
\end{equation}
Here the terms $A_{i}$ are referred to as \textquotedblleft drift
coefficients\textquotedblright\ and the terms $B_{ij}$ as \textquotedblleft
diffusion coefficients.\textquotedblright\ \ \ In work in 1924, Van
Vleck\cite{VV} carried out separate calculations for the mean change in the
mechanical action variables $J_{i}$ due to emission of radiation by the charge
in a small time $\delta t,$ the mean change in the action variables due to
absorption of radiation by the charge during $\delta t,$ and the mean squares
of the changes in the action variables during the time $\delta t.$ \ Marshall
and Claverie\cite{MC}\cite{M-Br}\cite{MC-Br} gave the average loss in the
drift coefficients $A_{i},$ and combined the average gain terms together with
the mean-square gain into the diffusion coefficients $B_{ij}$. \ Furthermore
they showed that the drift coefficients and diffusion coefficients could be
calculated from the unperturbed motion of the particle as
\begin{equation}
A_{i}=%
{\textstyle\sum\nolimits_{j}}
\left\langle \frac{2e^{2}}{3c^{3}}\dddot{x}_{j}\frac{\partial J_{i}}{\partial
p_{j}}\right\rangle =%
{\textstyle\sum\nolimits_{j}}
\left\langle \frac{2e^{2}}{3c^{3}}\dddot{x}_{j}\frac{\partial x_{j}}{\partial
w_{i}}\right\rangle
\end{equation}
and
\begin{align}
B_{ij}  &  =\frac{e^{2}}{2}%
{\textstyle\int\nolimits_{-\infty}^{\infty}}
B_{E}(\tau)%
{\textstyle\sum\nolimits_{k}}
\left\langle \frac{\partial J_{i}}{\partial p_{k}}(t)\frac{\partial J_{i}%
}{\partial p_{k}}(t+\tau)\right\rangle d\tau\nonumber\\
&  =\frac{e^{2}}{2}%
{\textstyle\int\nolimits_{-\infty}^{\infty}}
B_{E}(\tau)%
{\textstyle\sum\nolimits_{k}}
\left\langle \frac{\partial x_{k}}{\partial w_{i}}(t)\frac{\partial x_{k}%
}{\partial w_{j}}(t+\tau)\right\rangle d\tau,
\end{align}
where%
\begin{equation}
\delta_{ij}B_{E}\left(  \tau\right)  =\left\langle E_{i}\left(  0,t\right)
E_{j}\left(  0,t+\tau\right)  \right\rangle ,
\end{equation}
the brackets refer to averaging over time $t,$ and use has been made of the
relationship
\begin{equation}
\frac{\partial J_{i}}{\partial p_{j}}=\frac{\partial x_{j}}{\partial w_{i}}.
\label{canon}%
\end{equation}
This last relationship follows from the invariance of the Poisson bracket of
$x_{j}$ and $J_{i}$ under a canonical transformation from the $x_{i},p_{j}$
variables to the $w_{i},J_{j}$ variables.

In the present article, we will be interested in only the equilibrium
probability distribution $P(J_{1},J_{2},J_{3})$ with no time dependence, and
in situations involving radiation balance where there is no change in the
radiation spectrum, and therefore no energy current within the mechanical
system carrying energy from one frequency to another. \ In this case, the
Fokker-Planck equation in Eq. (\ref{FP}) becomes the condition for each index
$i$ that\cite{MC}%
\begin{equation}
0=-A_{i}P(J_{1},J_{2},J_{3})+%
{\textstyle\sum\nolimits_{j}}
B_{ij}\frac{\partial}{\partial J_{j}}P(J_{1},J_{2},J_{3}). \label{detbal}%
\end{equation}

In 1980, Marshall and Claverie\cite{MC} applied their understanding of
quasi-Markov processes to the classical hydrogen atom in classical
electromagnetic zero-point radiation. \ Our equation (\ref{detbal})
corresponds to their Eq. (5.2). \ They concluded that radiation balance did
not exist for this system, but rather the scattering by the orbiting electron
in the Coulomb potential was pushing the radiation spectrum away from the
Lorentz-invariant zero-point form over toward some other spectrum. \ Later
Claverie and Soto indicated that the hydrogen system was
self-ionizing.\cite{self}\ 

\section{Hydrogen System}

\subsection{Sensitive Connection to the Random Radiation Spectrum}

Although in 1965, Marshall showed that the spectrum of classical
electromagnetic zero-point radiation was Lorentz invariant,\cite{M65} the
physicists working with classical zero-point radiation in the 1960s and early
1970s did not appreciate the importance of using classical electromagnetic
systems which are consistent with relativistic behavior. \ Here we note that
the system of charged particles interacting through the Coulomb potential
(such as the classical hydrogen atom) can be extended to full classical
electrodynamics, and so provide a fully relativistic theory. \ Also, small
linear-oscillator systems can be regarded as relativistic in the limit of
small oscillation amplitude. \ 

Most of the successes for mechanical systems coupled to zero-point radiation
involve harmonic oscillator systems where the mechanical system has a
fundamental mechanical frequency $\omega_{0}$ and interacts through the dipole
approximation with radiation only at this same frequency $\omega_{0}$. \ On
the other hand, the classical hydrogen atom has several properties which are
in striking contrast with the properties of a harmonic oscillator. \ For
example, the classical hydrogen atom has no natural mechanical oscillation
frequency for the system. \ Indeed, a particle in a Coulomb potential allows
all frequencies with no limits on frequency, $0<\omega<\infty.$

Furthermore, the harmonic oscillator and the hydrogen atom are very different
in the appearance of time Fourier components associated with different
frequencies of oscillation. \ The purely harmonic oscillator has an electric
\textit{dipole} moment at the natural frequency $\omega_{0}$ of the oscillator
and at no other frequency. \ The mechanical motion is purely sinusoidal.
\ Even for the case of a charged particle in an isotropic
\textit{three-dimensional} harmonic potential with natural frequency
$\omega_{0}$ (where elliptical orbits appear), the small-amplitude motion is
equivalent to three independent one-dimensional linear oscillations, each
oscillation having an electric dipole moment oscillating at $\omega_{0}$ only;
there are no dipole moments at any higher multiples of $\omega_{0}$. \ In
complete contrast with this linear oscillator situation, a nonrelativistic
elliptical electron orbit in a Coulomb potential with eccentricity
$\epsilon>0~$has time Fourier components of velocity, not just at the
frequency $\omega$ of the orbital motion, but at all multiples of $\omega$.
\ Thus even in the most basic approximation, the \textit{dipole}
approximation, the elliptical nonrelativistic particle orbit (with the Coulomb
source at one focus) is connected to the electromagnetic radiation at not just
one frequency $\omega$ but at an infinite number of frequencies $n\omega$ for
all integers $n$. \ Thus a charged particle in a Coulomb potential is a
\textit{nonlinear} mechanical system where the mechanical motion alone will
determine the associated equilibrium spectrum of random radiation. \ The
\ classical hydrogen atom is an enormously sensitive system in connection with
radiation equilibrium since questions of radiation balance exist already in
the dipole approximation. \ 

We should emphasize that the one successful \textit{relativistic} scattering
calculation\cite{B2018} involves entirely different aspects from these
nonlinear systems. \ In complete contrast with nonlinear systems (such as
nonlinear oscillators or hydrogen), the successful relativistic calculation
involves a harmonic oscillator with just one time Fourier component of
velocity at the one fundamental oscillator frequency $\omega_{0}$. \ The
connection to the overtone frequencies is made, not through any time Fourier
components at higher frequencies (which are nonexistent), but rather through
the higher electric radiation\textit{\ multipole }moments (beyond the electric
\textit{dipole} moment, such as quadrupole, etc.) which are associated with
the finite amplitude of the oscillator motion and which connect the motion to
the full space-and-time dependence of the radiation field $\mathbf{E}\left(
\mathbf{r},t\right)  $.\ \ This relativistic calculation emphasizes the
importance of relativity but gives no information about the hydrogen system. \ 

For nonlinear oscillators treated in the dipole approximation, the
Fokker-Planck analysis developed by Marshall and Claverie makes clear that it
is the \textit{unperturbed mechanical} motion which determines the spectrum of
random radiation with which the oscillator is in equilibrium. \ The
relativistic nature of electromagnetic waves does not appear in the dipole
approximation which invokes $e\mathbf{E}\left(  0,t\right)  $ for the random
force nor in the term in $\mathbf{\dddot{r}}$ for radiation-damping.
\ Accordingly, any aspect of relativity must arise from the mechanical motion
of the charged particle. \ For the classical hydrogen atom, it is the time
Fourier components for the orbiting electron which will determine the spectrum
of radiation equilibrium. \ However, only the time Fourier components for
\textit{nonrelativistic} elliptical particle motions in a Coulomb potential
appear in the literature. \ Landau and Lifshitz give a calculation of the time
Fourier components of a nonrelativistic particle moving in a Coulomb
elliptical orbit,\cite{LL} and these dipole moments were then used in the
radiation-balance analysis of Marshall and Claverie. \ Because the time
Fourier components of the electromagnetic \textit{relativistic} particle
motion are different from those of the \textit{nonrelativistic} elliptical
orbital motion, the possibility arises that a treatment consistent with
relativistic behavior may give rise to radiation balance whereas the
completely nonrelativistic treatment does not.

\subsection{Qualitative Suggestion for Equilibrium}

The classical hydrogen atom in classical zero-point radiation has attracted
attention at least since 1975. \ In a review article published that year, it
was suggested that there was hope that a classical hydrogen atom in zero-point
radiation would have the same size as suggested by quantum theory.\cite{B1975}
It was noted\cite{B800} that the electron $e$ in a nonrelativistic circular
orbit where $\omega=\left[  e^{2}/\left(  mr^{3}\right)  \right]  ^{1/2}$
would emit radiation, and so lose nonrelativistic energy $\mathcal{E}$ at a
rate
\begin{equation}
\frac{d\mathcal{E}_{loss}}{dt}=\frac{2e^{2}}{3c^{3}}\omega^{4}r^{2}%
=\frac{2e^{6}}{3m^{2}c^{3}r^{4}}. \label{Uloss}%
\end{equation}
while the nonrelativistic energy gain from zero-point radiation was
\textit{estimated} as
\begin{equation}
\frac{d\mathcal{E}_{gain}}{dt}=\frac{e^{2}\hbar\omega^{3}}{2mc^{3}}%
=\frac{e^{5}\hbar}{2m^{5/2}c^{3}r^{9/2}} \label{Ugain}%
\end{equation}
by using the calculation for a rotating electric dipole in zero-point
radiation. \ Then energy balance required that $m\omega r^{2}=(3/4)\hbar$
which (except for the unreliable factor of $3/4)$ agrees with Bohr's condition
$m\omega r^{2}=\hbar$ for the hydrogen ground state in quantum theory.
$\ \ $In this qualitative analysis, there is no suggestion that relativistic
behavior for the particle motion may play a role in the equilibrium situation.

\subsection{\ Action-Angle Variables for Hydrogen}

The action-angle variables for a particle in a Coulomb potential have been
given in several books. \ The action-angle variables for nonrelativistic
hydrogen appear in Goldstein's textbook of classical mechanics;\cite{G} the
action-angle variables for both nonrelativistic and relativistic hydrogen are
given in Born's monograph, \textit{Mechanics of the Atom}.\cite{Born52} \ Only
the relativistic situation corresponds to electrodynamics. \ The component of
relativistic angular momentum along the $z$-axis is $J_{3}=J_{\phi}%
=m_{0}\gamma\left(  r\sin\theta\right)  ^{2}\dot{\phi}$ where $\gamma=\left(
1-v^{2}/c^{2}\right)  ^{-1/2}$, and the magnitude of the total angular
momentum is denote as $J_{2},$ $J_{2}=J_{\theta}+J_{\phi}.$ The relativistic
energy $U(J_{r},J_{\theta},J_{\phi})$\ (including the rest energy $m_{0}c^{2}%
$) is
\begin{equation}
U(J_{r},J_{\theta},J_{\phi})=m_{0}c^{2}\left(  1+\left[  \frac{e^{2}/c}%
{J_{r}+\sqrt{\left(  J_{\theta}+J_{\phi}\right)  ^{2}-(e^{2}/c)^{2}}}\right]
^{2}\right)  ^{-1/2}. \label{UJrJ2}%
\end{equation}
By taking derivatives of the Hamiltonian with respect to the action variables,
we can obtain the corresponding relativistic (angular) frequencies%

\begin{equation}
\omega_{r}=\frac{m_{0}c^{2}\left(  e^{2}/c\right)  ^{2}}{\left[  \left(
e^{2}/c\right)  ^{2}+\left(  J_{r}+\sqrt{J_{2}^{2}-\left(  e^{2}/c\right)
^{2}}\right)  ^{2}\right]  ^{3/2}}%
\end{equation}
and%
\begin{equation}
\omega_{\theta}=\omega_{\phi}=\omega_{2}=\frac{m_{0}c^{2}\left(
e^{2}/c\right)  ^{2}}{\sqrt{1-\left[  e^{2}/\left(  J_{2}c\right)  \right]
^{2}}\left[  \left(  e^{2}/c\right)  ^{2}+\left(  J_{r}+\sqrt{J_{2}%
^{2}-\left(  e^{2}/c\right)  ^{2}}\right)  ^{2}\right]  ^{3/2}}.
\end{equation}
We note that the relativistic expressions for $\omega_{r}$ and $\omega_{2}$
are not equal, rather $\omega_{r}=\Gamma\omega_{2}$ where there is a relative
factor of%
\[
\Gamma=\sqrt{1-\left[  e^{2}/\left(  J_{2}c\right)  \right]  ^{2}}.
\]

The relativistic orbital equation for a particle in a Coulomb potential is%
\begin{equation}
r=\frac{a\left(  1-\epsilon^{2}\right)  }{1+\epsilon\cos\left(  \Gamma
\phi\right)  } \label{rrell}%
\end{equation}
where the length $a$ depends only upon the energy $U$
\begin{equation}
a=\frac{e^{2}U}{\left(  m_{0}c^{2}\right)  ^{2}-U^{2}},
\end{equation}
and not upon the eccentricity $\epsilon$%
\begin{equation}
\epsilon=\sqrt{1-\frac{1}{\left[  1+J_{r}/\sqrt{J_{2}^{2}-\left(
e^{2}/c\right)  ^{2}}\right]  ^{2}}}=\sqrt{1-\frac{1}{\left[  1+J_{r}/\left(
J_{2}\Gamma\right)  \right]  ^{2}}}. \label{eccs}%
\end{equation}
The connection between the angle $\phi\left(  t\right)  $ and the time $t$
depends on the implicit relationship%
\begin{equation}
t=\int_{0}^{\phi}d\phi^{\prime}\frac{r^{2}[U+e^{2}/r]}{c^{2}J_{2}}=\left[
1+\frac{\left(  1-\epsilon^{2}\right)  }{\left[  \left(  m_{0}c^{2}/U\right)
^{2}-1\right]  }\right]  ^{-1/2}\int_{0}^{\phi}\frac{d\phi^{\prime}}{c}\left(
\frac{r^{2}U}{e^{2}}+r\right)  . \label{phioft}%
\end{equation}
\ 

\subsection{The Need to Compare the Relativistic and Nonrelativistic
Calculations}

All the calculation to date on the classical hydrogen atom in zero-point
radiation have been made in the nonrelativistic approximation. \ Thus the
suggestive qualitative calculation above\cite{B1975} in Eqs. (\ref{Uloss}) and
(\ref{Ugain}), the work of Marshall and Claverie,\cite{MC} and the simulations
of Cole and Zou\cite{CZ} have all started with a \textit{nonrelativistic}
charged particle in a Coulomb potential. \ However, it was emphasized in 2004
that \textit{relativistic} Coulomb orbits can be \textit{qualitatively}
different from the familiar conic-section orbits of nonrelativistic
physics.\cite{B204} \ Knowledge of this contrast between the nonrelativistic
and relativistic orbits, taken together with the new relativistic scattering
calculation,\cite{B2018} emphasizes that relativity may be important for the
classical hydrogen atom. \ The nonrelativistic limit corresponds to
$e^{2}/\left(  J_{2}c\right)  <<1.$ \ The nonrelativistic limits of all the
quantities in Eqs. (\ref{UJrJ2}) through (\ref{phioft}) are obtained in this
large-$J_{2}$ limit. \ 

\subsection{Multiply-Periodic Relativistic Coulomb Orbits}

In the dipole approximation, radiation balance depends upon the time Fourier
components of the electric dipole moment calculated from the particle's
unperturbed orbital motion. \ The dipole moment for the classical charged
particle in a Coulomb potential is obtained by multiplying the charge $e$
times the vector displacement, $e\mathbf{r}\left(  t\right)  =e\widehat{i}%
x\left(  t\right)  +e\widehat{j}y\left(  t\right)  .$ \ Now for a
multiply-periodic system involving a central potential, such as the
relativistic classical hydrogen atom, the $x$- and $y$-components of planar
orbits can be expanded in the form%

\begin{align}
x\left(  t\right)   &  =%
{\textstyle\sum\nolimits_{k}}
D_{k}\left(  J_{r},J_{2}\right)  \exp\left[  ikw_{r}+iw_{2}\right] \nonumber\\
&  =%
{\textstyle\sum\nolimits_{k}}
D_{k}\left(  J_{r},J_{2}\right)  \exp\left[  ik\left(  \omega_{r}t+\beta
_{r}\right)  +i\left(  \omega_{2}t+\beta2\right)  \right] \nonumber\\
&  =%
{\textstyle\sum\nolimits_{k}}
x_{k}\left(  J_{r},J_{2}\right)  \exp\left[  i\left(  1+k\Gamma\right)
\omega_{2}t\right]  \label{xtsum}%
\end{align}
where (up to a factor of $e$) the coefficients $x_{k}(J_{r},J_{2})$ will
provide the time Fourier components of the electric dipole moment oscillating
with angular frequency $\omega_{2}+k\omega_{r}=$ \ $\left(  1+k\Gamma\right)
\omega_{2}$.

\subsection{Periodic Nonrelativistic Coulomb Orbits}

Although the particle motion of a charged particle in a Coulomb potential in
electromagnetism is \textit{multiply periodic}, the motion of a
nonrelativistic particle in a Coulomb potential is \textit{periodic}. In
contrast with the time Fourier components of the relativistic orbital motion
$x_{k}$ given in Eq. (\ref{xtsum}), the time Fourier components $x_{n}^{NR}$
of the nonrelativistic elliptical orbital motion of the electron in the
Coulomb potential appear in the expansion%
\begin{align}
x_{NR}\left(  t\right)   &  =%
{\textstyle\sum\nolimits_{k}}
D_{NRk}\left(  J_{r},J_{2}\right)  \exp\left[  ikw_{r}+iw_{2}\right]
\nonumber\\
&  =%
{\textstyle\sum\nolimits_{k}}
D_{NRk}\left(  J_{r},J_{2}\right)  \exp\left[  ik\left(  \omega_{NR}%
t+\beta_{r}\right)  +i\left(  \omega_{NR}t+\beta_{2}\right)  \right]
\nonumber\\
&  =%
{\textstyle\sum\nolimits_{n}}
x_{n}^{NR}\exp\left[  in\omega_{NR}t\right]  , \label{xtNRsum}%
\end{align}
since in the nonrelativistic limit there is only one frequency involved,
$\omega_{NR}=\omega_{r}=\omega_{2}$. \ In the nonrelativistic limit (which
sends $\Gamma\rightarrow1$ since $c\rightarrow\infty$), each term in Eq.
(\ref{xtsum}) makes a contribution to one of the terms in Eq. (\ref{xtNRsum});
however, a single term in Eq. (\ref{xtNRsum}) may receive contributions from
more than one term in Eq. (\ref{xtsum}).

\subsection{Descriptions of Hydrogen Require Time Averaging}

Attempts to understand both the stability and radiation balance for the
hydrogen atom involve calculations over extended periods of time. \ Numerical
simulations follow the electron motion over many orbits, and the generalized
Fokker-Planck analysis (sketched above) requires time averaging over the
unperturbed particle motion in order to compute the drift and diffusion
coefficients. \ However, there is a crucial contrast between the relativistic
and nonrelativistic particle motion in a hydrogen atom. \ The relativistic
motion involves a rosette motion of a rotating elliptical orbit whereas the
nonrelativistic motion involves an elliptical orbit which is fixed in
orientation and is not rotating. \ On account of this contrast in fundamental
orbital motion, the relativistic and nonrelativistic time Fourier expansions
are different, and calculational results may depend upon whether the orbital
motion over an extended time is evaluated before or after taking the
nonrelativistic limit. \ 

If we first average over the rosette motion of the relativistic orbit, then
the time averages of $x^{2}$ and $y^{2}$ are identical, and when now taking
the nonrelativistic limit, this agreement between the time averages of $x^{2}$
and $y^{2}$ will persist. \ On the other hand, if we first take the
nonrelativistic limit for the particle motion, then the relativistic rosette
motion reduces to a nonrelativistic non-rotating ellipse whose orientation in
space is fixed. \ Let's assume (following Landau and Lifshitz) that the major
axis of the ellipse is along the $x$-axis and the minor axis along the
$y$-axis. \ Then for any ellipse with non-zero eccentricity $\epsilon>0,$ the
time average value of $x^{2}$ will be different from the time average of
$y^{2}$, since the squares of the semimajor and semiminor axes differ as
$a^{2}$ and $a^{2}\left(  1-\epsilon^{2}\right)  $. \ As the eccentricity
$\epsilon$ becomes larger, the difference between the time averages for
$x^{2}$ and $y^{2}$ will become larger, and hence the departure from the
relativistic situation becomes larger. \ 

The relativistic distinction between $\omega_{r}$ and $\omega_{2},$ which
leads to rosette motion, may well be important in connection with numerical
simulations involving energy absorption from random radiation. \ In the
nonrelativistic approximation which equates $\omega_{r}$ and $\omega_{2}$, the
unperturbed motion follows the same spatial path repeatedly, and so the
particle can pick up significant energy from the higher harmonics. \ On the
other hand, in the relativistic description, the spatial path is not repeated
but rather shifts in orientation and velocity, and so the particle is less
likely to pick up energy from the higher radiation harmonics which involve
higher frequencies and smaller wavelengths. \ This distinction may be
important in the numerical calculations (such as those of Nieuwenhuizen and
Liska\cite{NL}\cite{NL2} where self-ionization is found in orbits of high eccentricity.

\subsection{Dipole Approximation, Eccentricity, and Radiation Balance}

The use of the dipole approximation for the radiation emission term and the
radiation-forcing term in Eq. (\ref{Lang}) implies that the radiation speed is
vastly larger than any particle speed; there is no actual comparison between
the two. \ In the first Bohr orbit of old quantum theory, the electron speed
is nonrelativistic, and it is usually assumed that nonrelativistic
approximations are valid in any treatment of the hydrogen atom. \ However, the
comments above regarding the time Fourier coefficients of the relativistic
versus nonrelativistic orbital motions must make us cautious.

The use of the dipole approximation for the interaction between radiation and
matter means that radiation balance depends entirely upon the time Fourier
components of the unperturbed particle motion, and therefore upon the
eccentricity $\epsilon$. \ If the orbit is circular, $\epsilon=0$, then there
is exactly one coefficient $x_{1}(J_{2})$ at frequency $\omega_{2}$ in the
time Fourier expansion for the particle motion $x\left(  t\right)  $ in Eq.
(\ref{xtsum}). \ In this case, the mechanical motion interacts with the
radiation at this one frequency and at no other frequency; at a single time
$t,$ there is no transfer of energy between radiation at two different
frequencies. \ However, if the eccentricity of the particle orbit is non-zero,
$\epsilon>0$, then there are nonvanishing coefficients $x_{k}\left(
J_{r},J_{2}\right)  $ at several frequencies, and, at a single time $t,~$the
orbital motion of the charged particle can indeed transfer energy among
different radiation frequencies. \ For example, through first order in
$\epsilon,$ the orbital motion has time Fourier components at frequencies
$\omega_{2}+\omega_{r}$ and $\omega_{2}-\omega_{r},$ and so, at a single time
$t,$ the orbital motion can transfer energy between these two radiation
frequencies. \ As the eccentricity $\epsilon$ increases toward $\epsilon
\rightarrow1$, more and more time Fourier coefficients $x_{k}\left(
J_{r},J_{2}\right)  $\ will become important, and the mechanical motion can
transfer energy among many radiation frequencies. \ However, we will limit out
attention to the simplest case. The moment that more than one frequency is
involved, it is possible to test for radiation balance because the mechanical
system can transfer energy between the radiation modes at these differing
frequencies. \ Thus we will limit our calculations to particle motions which
are nearly circular and where the eccentricity $\epsilon~$is small. \ Our
calculations will hold through first order in $\epsilon$ for the
time-Fourier-coefficient expansions for $x\left(  t\right)  $ and $y\left(
t\right)  $, and through order $\epsilon^{2}$ for the squares of the particle
coordinates. \ 

\subsection{Failure of the General Calculation in Marshall and Claverie's
Analysis}

The limitation of our calculations to only first-order-in-$\epsilon$
departures from circular motion corresponds to the situation where the
nonrelativistic approximation is consistent with the accurate relativistic
analysis. \ We will not consider the enormously complex general problem of
radiation equilibrium at higher orders in $\epsilon$. \ Indeed, if we examine
equations (5.24) and (5.25) in Marshall and Claverie's work but stop at the
lowest order in $\epsilon$, then their criterion for radiation balance is
satisfied. \ However, Marshall and Claverie regarded their calculations as
valid to all orders in $\epsilon$, and such regard is erroneous. \ 

The failure of Marshall and Claverie's calculations goes back to a failure in
Landau and Lifshitz's \textit{Classical Theory of Fields}.\cite{LL} \ The
mathematics in the textbook is not in question. \ Rather, the error is in the
physical assumptions. \ Landau and Lifshitz assume that the nonrelativistic
limits of the time Fourier components of the relativistic particle motion are
the same as the time Fourier components of the nonrelativistic approximation
to the orbital motion. Landau and Lifshitz assume that the \textit{multiply
periodic} expansion for a relativistic charged particle motion in a Coulomb
potential given in our Eq. (\ref{xtsum}) can be represented in the
nonrelativistic limit by the \textit{periodic} expansion in Eq. (\ref{xtNRsum}%
). Such assumptions are not valid for the electrodynamics of a charge particle
in a Coulomb potential where the motion is necessarily relativistic. \ Thus
the time Fourier components $x_{k},y_{k}$ of the relativistic motion \ are
equal except for a relative phase corresponding to changing a cosine time
dependence into a sinusoidal time dependence. \ However, in the periodic
nonrelativistic elliptical motion assumed by Landau and Lifshitz, the time
Fourier components $x_{n}^{NR},y_{n}^{NR}$ are quite different corresponding
to expanding the semimajor and semiminor axes of an ellipse. \ In their
calculations, Landau and Lifshitz assume that the semimajor axis of the
stationary nonrelativistic ellipse is along the $x$-axis, and give the time
Fourier expansion coefficients following from (\ref{xtNRsum}) in terms of
Bessel functions $J_{n}(n\epsilon)$ and derivatives $J_{n}^{\prime}\left(
n\epsilon\right)  ,$ as%

\begin{equation}
x_{n}^{NR}=\frac{a}{n}J_{n}^{\prime}(n\epsilon)=\frac{a}{2}\left(
\frac{n\epsilon}{2}\right)  ^{n-1}+... \label{LLxn}%
\end{equation}
and%
\begin{equation}
y_{n}^{NR}=\frac{a\sqrt{1-\epsilon^{2}}}{n\epsilon}J_{n}\left(  n\epsilon
\right)  =\frac{a\sqrt{1-\epsilon^{2}}}{2}\left(  \frac{n\epsilon}{2}\right)
^{n-1}+... \label{LLyn}%
\end{equation}
Here, as in the later parts of Marshall and Claverie's work, the index $n$
follows the notation of Landau and Lifshitz rather that the earlier summation
notation given as $k$ in Eq. (\ref{xtNRsum}). \ These time Fourier expansion
coefficients start with $n=1.$ Use of Landau and Lifshitz's expressions for
$x_{n}^{NR}$ and $y_{n}^{NR}$ \textit{through first order in }$\epsilon$ leads
to the same results which we will present below. \ Indeed, only the first two
terms (corresponding to $n=1$ and $n=2)$ are given accurately through first
order in $\epsilon.$ \ In order $\epsilon^{2}$ and beyond, there is
disagreement with electrodynamics because of questions of the order of time
Fourier evaluations and nonrelativistic limits.\ \ 

\section{Radiation Balance for Hydrogen Orbits Through First Order in
Eccentricity}

\subsection{Calculation of Time Fourier Components Through First Order in
Eccentricity}

In the present article, we will consider the question of radiation balance
only through first order in the orbital eccentricity. \ We need to substitute
the relativistic orbital relations into \ \
\begin{equation}
x(t)=r\left[  \phi\left(  t\right)  \right]  \cos\phi(t),\label{xrphi}%
\end{equation}%
\begin{equation}
y\left(  t\right)  =r\left[  \phi\left(  t\right)  \right]  \sin\phi\left(
t\right)  .\label{yrphi}%
\end{equation}
Expanding the expression (\ref{rrell}) for $r(\phi)$ through first order in
$\epsilon$ gives%
\begin{equation}
r=a-\epsilon a\cos\Gamma\phi.\label{rphi}%
\end{equation}
Squaring this expression gives through order $\epsilon,$%
\begin{equation}
r^{2}=a^{2}-\epsilon2a^{2}\cos\Gamma\phi.\label{rsq1}%
\end{equation}
Substituting Eq. (\ref{rphi}) into Eqs. (\ref{xrphi}) and (\ref{yrphi}) and
using trigonometric identities, we have%
\begin{equation}
x=r\cos\phi=a\cos\phi-\frac{\epsilon a}{2}\left\{  \cos\left[  (1+\Gamma
)\phi\right]  +\cos\left[  (1-\Gamma)\phi\right]  \right\}  \label{xrrphi2}%
\end{equation}
and%
\begin{equation}
y=r\sin\phi=a\sin\phi-\frac{\epsilon a}{2}\left\{  \sin\left[  (1+\Gamma
)\phi\right]  +\sin\left[  (1-\Gamma)\phi\right]  \right\}  .\label{yrphi2}%
\end{equation}
We now need $\phi\left(  t\right)  $ through first order in $\epsilon$. \ We
use Eq. (\ref{phioft}) to obtain%
\begin{align}
t &  =\int_{0}^{\phi}d\phi^{\prime}\frac{r^{2}[U+e^{2}/r]}{c^{2}J_{2}}%
=\frac{U}{c^{2}J_{2}}\int_{0}^{\phi}d\phi^{\prime}r^{2}+\frac{e^{2}}%
{c^{2}J_{2}}\int_{0}^{\phi}d\phi^{\prime}r\nonumber\\
&  =\frac{U}{c^{2}J_{2}}\int_{0}^{\phi}d\phi^{\prime}\left\{  a^{2}%
-\epsilon2a^{2}\cos\Gamma\phi\right\}  +\frac{e^{2}}{c^{2}J_{2}}\int_{0}%
^{\phi}d\phi^{\prime}\left\{  a-\epsilon a\cos\Gamma\phi\right\}  \nonumber\\
&  =\left(  \frac{U}{c^{2}J_{2}}a^{2}+\frac{e^{2}}{c^{2}J_{2}}a\right)
\phi-\epsilon\left(  \frac{2a^{2}U}{c^{2}J_{2}\Gamma}+\frac{e^{2}a}{c^{2}%
J_{2}\Gamma}\right)  \sin\Gamma\phi.\label{tomB}%
\end{align}
We recognize%
\begin{equation}
\frac{1}{\omega_{2}}=\frac{a}{c^{2}J_{2}}\left(  aU+e^{2}\right)  ,
\end{equation}
and, for simplicity, we define%
\begin{equation}
F=\frac{2aU+e^{2}}{aU+e^{2}}=2-\left(  \frac{e^{2}}{J_{2}c}\right)  ^{2}.
\end{equation}
Then Eq. (\ref{tomB}) becomes%
\begin{equation}
\omega_{2}t=\phi-\epsilon\frac{F}{\Gamma}\sin\Gamma\phi,
\end{equation}
and, through first order in $\epsilon,$%
\begin{equation}
\phi=\omega_{2}t+\epsilon\frac{F}{\Gamma}\sin\omega_{2}t.
\end{equation}
Substituting this result for $\phi\left(  t\right)  $ into Eqs. (\ref{xrrphi2}%
) and (\ref{yrphi2}), we find through first order in $\epsilon$%
\begin{equation}
x=a\cos\omega_{2}t+\epsilon a\left(  -\frac{1}{2}+\frac{F}{2\Gamma}\right)
\cos\left[  \left(  1+\Gamma\right)  \omega_{2}t\right]  -\epsilon a\left(
\frac{1}{2}+\frac{F}{2\Gamma}\right)  \cos\left[  \left(  1-\Gamma\right)
\omega_{2}t\right]  \label{xrel}%
\end{equation}
and%
\begin{equation}
y=a\sin\omega_{2}t+\epsilon a\left(  -\frac{1}{2}+\frac{F}{2\Gamma}\right)
\sin\left[  \left(  1+\Gamma\right)  \omega_{2}t\right]  -\epsilon a\left(
\frac{1}{2}+\frac{F}{2\Gamma}\right)  \sin\left[  \left(  1-\Gamma\right)
\omega_{2}t\right]  .\label{yrel}%
\end{equation}
Here we have the time Fourier expansion through first order in $\epsilon$ for
the relativistic particle motion. \ Note that for these expressions which
assume relativistic motion, we have exactly the same coefficients for the
corresponding terms in $x$ and in $y$, which involve frequencies $\omega
_{2},\omega_{2}\pm\omega_{r}=\left(  1\pm\Gamma\right)  \omega_{2}.$

If we take the nonrelativistic limit of these expression, then $\Gamma
\rightarrow1$ and $F\rightarrow2,$ and we notice a contrast between the
expressions for $x_{NR}\left(  t\right)  $ and $y_{NR}\left(  t\right)  $
arising from $\cos\left[  \left(  1-\Gamma\right)  \omega_{2}t\right]
\rightarrow\cos0=1$ and $\sin\left[  \left(  1-\Gamma\right)  \omega
_{2}t\right]  \rightarrow\sin0=0.$ Thus the nonrelativistic limits of the time
Fourier expansions for the relativistic particle motion become%
\begin{equation}
x\rightarrow x_{NR}=-\frac{3\epsilon a}{2}+a\cos\omega_{NR}t+\frac{\epsilon
a}{2}\cos\left[  2\omega_{NR}t\right]  \label{xgoes}%
\end{equation}
and
\begin{equation}
y\rightarrow y_{NR}=a\sin\omega_{NR}t+\frac{\epsilon a}{2}\sin2\omega_{NR}t.
\label{ygoes}%
\end{equation}
These expressions give the time Fourier components through first order in
$\epsilon$ for $x_{NR}\left(  t\right)  $ and $y_{NR}\left(  t\right)  $ at
frequencies $\omega_{NR}$ and $2\omega_{NR}.$

Next we note the difference between the squares of the components depending
upon the orders of time averaging and nonrelativistic limit. \ If we time
average the squares of the relativistic expressions in Eqs. (\ref{xrel}) and
(\ref{yrel}), and subsequently take the nonrelativistic limit, we have equal
contributions for $\left\langle x^{2}\right\rangle $ and $\left\langle
y^{2}\right\rangle ,$
\begin{align}
\left\langle x^{2}\right\rangle  &  =\left\langle y^{2}\right\rangle =\frac
{1}{2}\left[  a^{2}+\epsilon^{2}a^{2}\left(  -\frac{1}{2}+\frac{F}{2\Gamma
}\right)  ^{2}+\epsilon^{2}a^{2}\left(  \frac{1}{2}+\frac{F}{2\Gamma}\right)
^{2}\right] \nonumber\\
&  \rightarrow\frac{1}{2}\left[  a^{2}+\epsilon^{2}a^{2}\left(  \frac{1}%
{2}\right)  ^{2}+\epsilon^{2}a^{2}\left(  \frac{3}{2}\right)  ^{2}\right]
=\frac{a^{2}}{2}+\frac{5}{4}\epsilon^{2}a^{2}.
\end{align}
On the other hand, if the nonrelativistic limit is taken as in Eqs.
(\ref{xgoes}) and (\ref{ygoes}) before the time averaging, we find that
$\left\langle x_{NR}^{2}\right\rangle $ and $\left\langle y_{NR}%
^{2}\right\rangle $ are quite different,
\begin{equation}
\left\langle x_{NR}^{2}\right\rangle =\left(  \frac{3\epsilon a}{2}\right)
^{2}+\frac{1}{2}\left[  a^{2}+\left(  \frac{\epsilon a}{2}\right)
^{2}\right]  =\frac{a^{2}}{2}+\frac{19}{8}\epsilon^{2}a^{2}%
\end{equation}
and
\begin{equation}
\left\langle y_{NR}^{2}\right\rangle =\frac{1}{2}\left[  a^{2}+\left(
\frac{\epsilon a}{2}\right)  ^{2}\right]  =\frac{a^{2}}{2}+\frac{1}{8}%
\epsilon^{2}a^{2}.
\end{equation}
When dealing with higher-order terms in the eccentricity $\epsilon$, the
nonrelativistic limit will involve terms beyond the simple relative constant
of $-3\epsilon a/2$ which appears between Eqs. (\ref{xgoes}) and
(\ref{ygoes}). \ For example, terms in $\epsilon^{2}$ in $x$ and $y$ will
introduce frequencies $\left(  1+2\Gamma\right)  \omega_{2}$ and $\left(
1-2\Gamma\right)  \omega_{2}.$ \ In the nonrelativistic limits, these
frequencies become $3\omega_{NR}$ and $\omega_{NR},$ so that the coefficients
of the terms in $\cos\omega_{NR}t$ and $\sin\omega_{NR}t$ will be altered.
\ With higher order terms in the nonrelativistic approximation, the time
Fourier expansions for $x$ and $y$ become increasingly different, and the
departure from valid electrodynamics increases.

Landau and Lifshitz give the time Fourier coefficients for the
\textit{velocity} of a nonrelativistic particle in a Coulomb potential, and so
they do not include the constant term which appears in our nonrelativistic
limit for $x$ given in Eq. (\ref{xgoes}). \ Because Landau and Lifshitz take
the limit over to nonrelativistic orbital motion before obtaining the time
Fourier coefficients of the particle motion, their time Fourier coefficients
correspond to electromagnetism only through first order in $\epsilon.$ \ 

\subsection{Obtaining the Fokker-Planck Equations}

From the time Fourier expansions in Eqs. (\ref{xrel}) and (\ref{yrel}), it is
clear that for the relativistic particle motion through order $\epsilon$, we
will have time Fourier components at frequencies $\omega_{2},$ $\left(
1+\Gamma\right)  \omega_{2}$, and $\left(  1-\Gamma\right)  \omega_{2}$.
\ Thus the relativistic electron in the Coulomb potential can transfer
energies among these frequencies. \ 

In order to take advantage of Marshall and Claverie's analysis,\cite{MC} we
will rewrite Eqs. (\ref{xrel}) and (\ref{yrel}) in terms of the angle
variables $w_{r}=\omega_{r}t+\beta_{r}$ and $w_{2}=\omega_{2}t+\beta_{2}$.
\ \ We have through terms first order in $\epsilon$%

\begin{equation}
x(t)=a\cos w_{2}+\epsilon a\left[  G\cos\left(  w_{2}+w_{r}\right)
-H\cos\left(  w_{2}-w_{r}\right)  \right]  ,\label{xaeps}%
\end{equation}
and%
\begin{equation}
y(t)=a\sin w_{2}+\epsilon a\left[  G\sin\left(  w_{2}+w_{r}\right)
-H\sin\left(  w_{2}-w_{r}\right)  \right]  ,\label{yaeps}%
\end{equation}
where
\begin{equation}
G=\frac{1}{2\Gamma}\left(  -\Gamma+F\right)  =\frac{1}{2\Gamma}\left(
-\Gamma+\frac{2aU+e^{2}}{aU+e^{2}}\right)
\end{equation}
and
\begin{equation}
H=\frac{1}{2\Gamma}\left(  \Gamma+F\right)  =\frac{1}{2\Gamma}\left(
\Gamma+\frac{2aU+e^{2}}{aU+e^{2}}\right)  .
\end{equation}

Now we want to evaluate Marshall and Claverie's quantities $A_{r},A_{2,}%
B_{rr},B_{r2},B_{22}$. \ We will need the quantity following from Eq.
(\ref{canon})
\begin{equation}
\left\langle \dddot{x}\left(  t\right)  \frac{\partial J_{r}}{\partial p_{x}%
}+\dddot{y}\left(  t\right)  \frac{\partial J_{r}}{\partial p_{y}%
}\right\rangle =\left\langle \dddot{x}\left(  t\right)  \frac{\partial
x}{\partial w_{r}}+\dddot{y}\left(  t\right)  \frac{\partial y}{\partial
w_{r}}\right\rangle .\label{avr}%
\end{equation}
We then evaluate $\dddot{x}\left(  t\right)  $ and $\partial x/\partial w_{r}$
from Eq. (\ref{xaeps}), and $\dddot{y}\left(  t\right)  $ and $\partial
y/\partial w_{r}$ from Eq. (\ref{yaeps}), giving%
\begin{align}
&  \left\langle \dddot{x}\left(  t\right)  \frac{\partial x}{\partial w_{r}%
}\right\rangle \nonumber\\
&  =\left\langle \left[  a\omega_{2}^{3}\sin w_{2}+\epsilon a\left[  G\left(
\omega_{2}+\omega_{r}\right)  ^{3}\sin\left(  w_{2}+w_{r}\right)  +H\left(
\omega_{2}-\omega_{r}\right)  ^{3}\sin\left(  w_{2}-w_{r}\right)  \right]
\right]  \right.  \nonumber\\
&  \left.  \times\left\{  0+\epsilon a\left[  -G\sin\left(  w_{2}%
+w_{r}\right)  -H\sin\left(  w_{2}-w_{r}\right)  \right]  \right\}
\right\rangle \label{avrr}%
\end{align}
since $\partial w_{r}/\partial t=\omega_{r}$ and $\partial w_{2}/\partial
t=\omega_{2}$, and there is a similar term in $\left\langle \dddot{y}\partial
y/\partial w_{r}\right\rangle .$ \ Here on time averaging, we have
$w_{r}=\omega_{r}t+\beta_{r}$ and $w_{2}=\omega_{2}t+\beta_{2}$ so that
\begin{equation}
\left\langle \sin^{2}w\right\rangle =\left\langle \cos^{2}w\right\rangle =1/2
\end{equation}
while%
\begin{equation}
\left\langle \sin w\cos w\right\rangle =0.
\end{equation}
Then combining Eqs. (\ref{avr}) and (\ref{avrr}), and the analogous terms in
$y,$ we have
\begin{align}
A_{r} &  =\frac{2e^{2}}{3c^{3}}\left\langle \dddot{x}\left(  t\right)
\frac{\partial J_{r}}{\partial p_{x}}+\dddot{y}\left(  t\right)
\frac{\partial J_{r}}{\partial p_{y}}\right\rangle =\frac{2e^{2}}{3c^{3}%
}\left\langle \dddot{x}\left(  t\right)  \frac{\partial x}{\partial w_{r}%
}+\dddot{y}\left(  t\right)  \frac{\partial y}{\partial w_{r}}\right\rangle
\nonumber\\
&  =-\frac{2e^{2}}{3c^{3}}\frac{\epsilon^{2}a^{2}}{2}\left[  G^{2}\left(
\omega_{2}+\omega_{r}\right)  ^{3}+H^{2}\left(  \omega_{2}-\omega_{r}\right)
^{3}\right]  .\label{ArR}%
\end{align}
Proceeding in a similar fashion, we obtain%
\begin{align}
A_{2} &  =\frac{2e^{2}}{3c^{3}}\left\langle \dddot{x}\left(  t\right)
\frac{\partial J_{2}}{\partial p_{x}}+\dddot{y}\left(  t\right)
\frac{\partial J_{2}}{\partial p_{y}}\right\rangle =\frac{2e^{2}}{3c^{3}%
}\left\langle \dddot{x}\left(  t\right)  \frac{\partial x}{\partial w_{2}%
}+\dddot{y}\left(  t\right)  \frac{\partial y}{\partial w_{2}}\right\rangle
\nonumber\\
&  =-\frac{2e^{2}}{3c^{3}}\left\{  a^{2}\omega_{2}^{3}+\frac{\epsilon^{2}%
a^{2}}{2}\left[  G^{2}\left(  \omega_{2}+\omega_{r}\right)  ^{3}+H^{2}\left(
\omega_{2}-\omega_{r}\right)  ^{3}\right]  \right\}  .\label{A2R}%
\end{align}

Next we need averages over the time $t$ involving%
\[
\left\langle \frac{\partial J_{r}}{\partial p_{x}}(t)\frac{\partial J_{r}%
}{\partial p_{x}}\left(  t+\tau\right)  +\frac{\partial J_{r}}{\partial p_{y}%
}(t)\frac{\partial J_{r}}{\partial p_{y}}\left(  t+\tau\right)  \right\rangle
.
\]
The term in $x$ involves
\begin{align}
&  \left\langle \frac{\partial J_{r}}{\partial p_{x}}(t)\frac{\partial J_{r}%
}{\partial p_{x}}\left(  t+\tau\right)  \right\rangle =\left\langle
\frac{\partial x}{\partial w_{r}}(t)\frac{\partial x}{\partial w_{r}}\left(
t+\tau\right)  \right\rangle \nonumber\\
&  =\left\langle \left\{  0+\epsilon a\left[  -G\sin\left(  w_{2}%
+w_{r}\right)  -H\sin\left(  w_{2}-w_{r}\right)  \right]  \right\}  \right.
\nonumber\\
&  \times\left.  \left\{  0+\epsilon a\left[  -G\sin\left(  w_{2}+w_{r}%
+\omega_{2}\tau+\omega_{r}\tau\right)  -H\sin\left(  w_{2}-w_{r}+\omega
_{2}\tau-\omega_{r}\tau\right)  \right]  \right\}  \right\rangle \nonumber\\
&  =\frac{\epsilon^{2}a^{2}}{2}\left[  G^{2}\cos\left(  \omega_{2}\tau
+\omega_{r}\tau\right)  +H^{2}\cos\left(  \omega_{2}\tau-\omega_{r}%
\tau\right)  \right]  ,\label{JpJp}%
\end{align}
with an analogous term in $y\left(  t\right)  $. \ Then we find%
\begin{align}
B_{rr} &  =\frac{e^{2}}{2}%
{\textstyle\int\nolimits_{-\infty}^{\infty}}
d\tau B_{E}\left(  \tau\right)  \left\langle \frac{\partial J_{r}}{\partial
p_{x}}(t)\frac{\partial J_{r}}{\partial p_{x}}\left(  t+\tau\right)
+\frac{\partial J_{r}}{\partial p_{y}}(t)\frac{\partial J_{r}}{\partial p_{y}%
}\left(  t+\tau\right)  \right\rangle \nonumber\\
&  =\frac{e^{2}}{2}\epsilon^{2}a^{2}%
{\textstyle\int\nolimits_{-\infty}^{\infty}}
d\tau B_{E}\left(  \tau\right)  \left[  G^{2}\cos\left(  \omega_{2}\tau
+\omega_{r}\tau\right)  +H^{2}\cos\left(  \omega_{2}\tau-\omega_{r}%
\tau\right)  \right]  \nonumber\\
&  =\epsilon^{2}a^{2}\frac{2e^{2}}{3c^{3}}\frac{\hbar}{2}\left\{  G^{2}\left(
\omega_{2}+\omega_{r}\right)  ^{3}+H^{2}\left(  \omega_{2}-\omega_{r}\right)
^{3}\right\}  ,\label{Brrzp}%
\end{align}
where we evaluated the time integrals as in Eqs. (\ref{SEE}) and (\ref{SEEzp})
and have used $S_{E\text{zp}}\left(  \omega\right)  $ as given in Eq.
(\ref{SEEzp}). In a similar fashion, we obtain
\begin{align}
B_{r2} &  =\frac{e^{2}}{2}%
{\textstyle\int\nolimits_{-\infty}^{\infty}}
d\tau B_{E}\left(  \tau\right)  \left\langle \frac{\partial J_{r}}{\partial
p_{x}}(t)\frac{\partial J_{2}}{\partial p_{x}}\left(  t+\tau\right)
+\frac{\partial J_{r}}{\partial p_{y}}(t)\frac{\partial J_{2}}{\partial p_{y}%
}\left(  t+\tau\right)  \right\rangle \nonumber\\
&  =\frac{e^{2}}{2}\epsilon^{2}a^{2}%
{\textstyle\int\nolimits_{-\infty}^{\infty}}
d\tau B_{E}\left(  \tau\right)  \left[  G^{2}\cos\left(  \omega_{2}\tau
+\omega_{r}\tau\right)  +H^{2}\cos\left(  \omega_{2}\tau-\omega_{r}%
\tau\right)  \right]  \nonumber\\
&  =\epsilon^{2}a^{2}\frac{2e^{2}}{3c^{3}}\frac{\hbar}{2}\left\{  G^{2}\left(
\omega_{2}+\omega_{r}\right)  ^{3}+H^{2}\left(  \omega_{2}-\omega_{r}\right)
^{3}\right\}  \label{Br2zp}%
\end{align}
and
\begin{align}
B_{22} &  =\frac{e^{2}}{2}%
{\textstyle\int\nolimits_{-\infty}^{\infty}}
d\tau B_{E}\left(  \tau\right)  \left\langle \frac{\partial J_{2}}{\partial
p_{x}}(t)\frac{\partial J_{2}}{\partial p_{x}}\left(  t+\tau\right)
+\frac{\partial J_{2}}{\partial p_{y}}(t)\frac{\partial J_{2}}{\partial p_{y}%
}\left(  t+\tau\right)  \right\rangle \nonumber\\
&  =\frac{e^{2}}{2}\left\{  a^{2}%
{\textstyle\int\nolimits_{-\infty}^{\infty}}
d\tau B_{E}\left(  \tau\right)  \cos\omega_{2}\tau\right.  \nonumber\\
&  +\left.  \epsilon^{2}a^{2}%
{\textstyle\int\nolimits_{-\infty}^{\infty}}
d\tau B_{E}\left(  \tau\right)  \left[  G^{2}\cos\left(  \omega_{2}\tau
+\omega_{r}\tau\right)  +H^{2}\cos\left(  \omega_{2}\tau-\omega_{r}%
\tau\right)  \right]  \right\}  \nonumber\\
&  =a^{2}\frac{2e^{2}}{3c^{3}}\frac{\hbar}{2}\omega_{2}^{3}+\epsilon^{2}%
a^{2}\frac{2e^{2}}{3c^{3}}\frac{\hbar}{2}\left[  G^{2}\left(  \omega
_{2}+\omega_{r}\right)  ^{3}+H^{2}\left(  \omega_{2}-\omega_{r}\right)
^{3}\right]  .\label{B22zp}%
\end{align}

\subsection{Using the Fokker-Planck Equations Associated with Radiation
Balance}

\subsubsection{Circular Orbits}

In the case of circular motion where the eccentricity vanishes, $\epsilon=0,$
the particle motion is connected to only one frequency, the fundamental
frequency $\omega_{2}$. \ In this case, radiation balance always holds since
at any time $t$ energy can be exchanged with only one frequency. \ In this
circular case, the action variable $J_{r}$ vanishes, and Marshall and
Claverie's coefficients simplify to $A_{r}=0,$ $B_{rr}=0,~B_{r2}=0,$
\begin{equation}
A_{2}=-\frac{2e^{2}}{3c^{3}}\left\{  a^{2}\omega_{2}^{3}\right\}  ,\text{
\ circular case.}%
\end{equation}
and%
\begin{equation}
B_{22}=\frac{2e^{2}}{3c^{3}}\frac{\hbar}{2}\left\{  a^{2}\omega_{2}%
^{3}\right\}  ,\text{ \ circular case.}%
\end{equation}
Then the non-vanishing equation for equilibrium requires%
\begin{align*}
0  &  =-A_{2}P(J_{2})+B_{22}\frac{dP(J_{2})}{dJ_{2}}\\
&  =\frac{2e^{2}}{3c^{3}}\left\{  a^{2}\omega_{2}^{3}\right\}  P(J_{2}%
)+\frac{2e^{2}}{3c^{3}}\frac{\hbar}{2}\left\{  a^{2}\omega_{2}^{3}\right\}
\frac{dP(J_{2})}{dJ_{2}}%
\end{align*}
with the solution
\begin{equation}
P\left(  J_{2}\right)  =const\times\exp\left[  -\frac{J_{2}}{\hbar/2}\right]
. \label{PJ2res}%
\end{equation}
The solution suggests that the average value of the action variable $J_{2}$ of
the mechanical Coulomb system is directly connected to Planck's constant
$\hbar$ appearing as the multiplicative scale for zero-point radiation.

\subsubsection{Orbits of Small Eccentricity}

For the situation of circular orbits, there is no question about radiation
balance, since the phase space distribution assures that the particle is in
equilibrium with the radiation at the single frequency $\omega_{2}.$ \ The
crucial question is the radiation balance when the non-zero eccentricity of
the orbit connects quantities at different frequencies such as $\omega_{2},$
$\omega_{2}+\omega_{r},$ and $\omega_{2}-\omega_{r}.$ \ 

Now using the full drift and diffusion coefficients in Eq. (\ref{ArR}%
)-(\ref{B22zp}), the first Fokker-Planck equation, labeled by the index $r,$
requires%
\begin{align}
0 &  =-A_{r}P(J_{r},J_{2})+B_{rr}\frac{dP(J_{r},J_{2})}{dJ_{r}}+B_{r2}%
\frac{dP(J_{r},J_{2})}{dJ_{2}}\nonumber\\
&  =\frac{2e^{2}}{3c^{3}}\epsilon^{2}a^{2}\left[  G^{2}\left(  \omega
_{2}+\omega_{r}\right)  ^{3}+H^{2}\left(  \omega_{2}-\omega_{r}\right)
^{3}\right]  P(J_{r},J_{2})\nonumber\\
&  +\frac{2e^{2}}{3c^{3}}\epsilon^{2}a^{2}\frac{\hbar}{2}\left\{  G^{2}\left(
\omega_{2}+\omega_{r}\right)  ^{3}+H^{2}\left(  \omega_{2}-\omega_{r}\right)
^{3}\right\}  \frac{dP(J_{r},J_{2})}{dJ_{r}}\nonumber\\
&  +\frac{2e^{2}}{3c^{3}}\epsilon^{2}a^{2}\frac{\hbar}{2}\left\{  G^{2}\left(
\omega_{2}+\omega_{r}\right)  ^{3}+H^{2}\left(  \omega_{2}-\omega_{r}\right)
^{3}\right\}  \frac{dP(J_{r},J_{2})}{dJ_{2}},\label{First}%
\end{align}
and the second, labeled by the index $2,$ requires
\begin{align}
0 &  =-A_{2}P(J_{r},J_{2})+B_{2r}\frac{dP(J_{r},J_{2})}{dJ_{r}}+B_{22}%
\frac{dP(J_{r},J_{2})}{dJ_{2}}\nonumber\\
&  =\frac{2e^{2}}{3c^{3}}\left\{  a^{2}\omega_{2}^{3}+\epsilon^{2}a^{2}\left[
G^{2}\left(  \omega_{2}+\omega_{r}\right)  ^{3}+H^{2}\left(  \omega_{2}%
-\omega_{r}\right)  ^{3}\right]  \right\}  P(J_{r},J_{2})\nonumber\\
&  +\frac{2e^{2}}{3c^{3}}\epsilon^{2}a^{2}\frac{\hbar}{2}\left\{  G^{2}\left(
\omega_{2}+\omega_{r}\right)  ^{3}+H^{2}\left(  \omega_{2}-\omega_{r}\right)
^{3}\right\}  \frac{dP(J_{r},J_{2})}{dJ_{r}}\nonumber\\
&  +\frac{2e^{2}}{3c^{3}}\frac{\hbar}{2}\left\{  a^{2}\omega_{2}^{3}%
+\epsilon^{2}a^{2}\left[  G^{2}\left(  \omega_{2}+\omega_{r}\right)
^{3}+H^{2}\left(  \omega_{2}-\omega_{r}\right)  ^{3}\right]  \right\}
\frac{dP(J_{r},J_{2})}{dJ_{2}}.\label{Second}%
\end{align}
However, by subtracting one equation from the other and cancelling common
factors, the two equations can be reduced to the two differential equations%
\begin{equation}
0=P(J_{r},J_{2})+\frac{\hbar}{2}\frac{dP(J_{r},J_{2})}{dJ_{2}}.\label{PJrJ2}%
\end{equation}
and
\begin{equation}
0=\frac{\hbar}{2}\frac{dP(J_{r},J_{2})}{dJ_{r}}.\label{EqPJr}%
\end{equation}
The last equation (\ref{EqPJr}) means that (through lowest nonvanishing order
in $\epsilon)$ the probability $P(J_{r},J_{2})$ does not depend upon $J_{r}$,
while the solution of (\ref{PJrJ2}) is the same as in Eq. (\ref{PJ2res}).
\ This phase space solution (\ref{PJ2res}) indeed works as a solution for both
equations (\ref{First}) and (\ref{Second}). \ Therefore radiation balance
indeed holds through lowest nonvanishing order in $\epsilon$. \ In this order
in $\epsilon,$ the lowest order distribution $P(J_{2})$ which gives radiation
balance at $\omega_{2}$ will also produce radiation balance at $\omega
_{2}+\omega_{r}$ and $\omega_{2}-\omega_{r}$ . \ 

\subsection{Marshall and Claverie's Analysis Through First Order in
Eccentricity}

In the analysis here, we have used the generalized Fokker-Planck equation of
Marshall and Claverie while taking time Fourier expansion coefficients
corresponding to the relativistic orbital motion with frequencies $\omega
_{2},$ $\omega_{2}+\omega_{r},$ and $\omega_{2}-\omega_{r}$. \ However,
through first order in the eccentricity $\epsilon,$ the relativistic time
Fourier coefficients go over to the nonrelativistic coefficients given by
Landau and Lifshitz in our Eqs. (\ref{LLxn}) and (\ref{LLyn}). \ Thus we can
take the nonrelativistic limits given in Eqs. (\ref{xgoes}) and (\ref{ygoes}),
and substitute these when calculating the drift and diffusion coefficients in
Eqs. (\ref{avr}) through (\ref{B22zp}). \ This nonrelativistic limit involves
the fundamental frequency $\omega_{2}\rightarrow\omega_{NR}$ of the
nonrelativistic orbital motion and its first overtone $\omega_{2}+\omega
_{r}\rightarrow2\omega_{NR}$. \ The calculations leading to the resulting
phase space in Eq. (\ref{PJ2res}) remain valid and radiation balance still
holds. \ Thus Marshall and Claverie's analysis gives radiation balance,
involving the particle velocity coefficients through lowest nonvanishing order
in the eccentricity $\epsilon.$ \ The failure of radiation balance claimed by
Marshall and Claverie involves higher powers in the eccentricity $\epsilon$
where the time Fourier components of the nonrelativistic orbit are not a valid
approximation to relativistic electrodynamics.

\section{Concluding Remarks}

\subsection{Unique Aspects of the Coulomb Potential and Zero-Point Radiation}

The Coulomb potential is part of relativistic electromagnetic theory as is
Lorentz-invariant zero-point radiation. The Coulomb potential $V\left(
r\right)  =-e^{2}/r$ is very special in its scaling\cite{B2007} which gives
hydrogen its dependence upon the one parameter of mass $m_{0},$ and leads to
the velocity in a circular orbit given by $v=e^{2}/J_{2}$ with no other
parameters involved. \ Zero-point radiation is scale invariant,\cite{B2007}
and combines with the Coulomb potential to give the phase space distribution
for circular orbits in Eq. (\ref{PJ2res}) which makes no reference to the
frequency of the circular orbital motion. \ It is this independence from the
frequency which allows the radiation balance which we have noted through the
lowest nonvanishing order in the eccentricity $\epsilon.$ \ 

Any isotropic potential (such as $V\left(  r\right)  =-\alpha/r^{n})$ will
have circular orbits which have dipole moments connecting the circular orbit
of radius $r$ to a single frequency of radiation $\omega\left(  r\right)  $.
\ Thus we should always be able to find an equilibrium phase space associated
with circular orbits and some isotropic spectrum $U_{rad}(\omega)$ of random
radiation. \ However, unlike Eq. (\ref{PJ2res}), these phase space
distributions for a general potential and spectrum will depend upon the
orbital frequency, and so will not give radiation balance for any distortion
of the circular orbital motion.

In addition to the situation of the Coulomb potential and zero-point radiation
treated above, there is one other special case where radiation balance holds.
\ Using the Marshall-Claverie analysis in the dipole approximation (sketched
above), it is easy to show that a multiply periodic mechanical system
(\ref{xtsum}) in the Rayleigh-Jeans spectrum $U_{rad}(\omega)=k_{B}T$ leads to
the Boltzmann distribution $P(J_{1},J_{2},J_{3})=const\times\exp\left[
-U\left(  J_{1},J_{2},J_{3}\right)  /\left(  k_{B}T\right)  \right]  $ for the
equilibrium phase space with radiation balance. \ This calculation was first
carried out and published in part by van Vleck in 1924. \ However, this
analysis makes no sense for the Coulomb potential where the Boltzmann
probability density cannot be normalized. \ The Coulomb potential of
electrodynamics is a very special potential connected to relativity. \ 

\subsection{Symmetries and Radiation Balance}

Our analysis has been very limited in scope. \ However, there are general
principles which encourage the belief that the physical ideas hold more
generally. \ The Coulomb potential is part of Lorentz-invariant classical
electrodynamics while classical electromagnetic zero-point radiation has a
Lorentz invariant spectrum. \ How can a relativistic scatterer, such as the
relativistic classical hydrogen atom, push the zero-point spectrum toward a
spectrum which is not Lorentz invariant? \ It seems natural to suggest that
the Lorentz invariance of the zero-point spectrum must be preserved by
relativistic Coulomb scattering, as demonstrated here through lowest
nonvanishing order in the orbital eccentricity.

\subsection{Speculation}

The existence of radiation balance in classical zero-point radiation will hide
the electron's radiation emission and also its energy absorption in the
classical hydrogen atom. \ Thus in equilibrium, an observer would not detect
net radiation emission or absorption in any frequency range if radiation
balance holds. \ One may speculate that the excited states of atoms correspond
to additional configurations at energies above the ground state where
radiation balance again exists at least approximately, and that the observed
radiation spectra arise from radiation emission by a classical charged
particle (electron) when moving from an exited configuration to a lower energy
configuration. \ What is involved is the interaction with random radiation in
a non-Markov process where relativity is crucial. \ If we use nonrelativistic
calculations, these calculations must involve valid nonrelativistic
approximations to the true relativistic situation in nature.

\subsection{Conclusion}

There are four significant messages from the present analysis of the classical
hydrogen atom in classical zero-point radiation. \ 1) In electromagnetism,
aspects of relativity may be important even for charged particle motions which
have low velocities. \ 2) Zero-point radiation gives equilibrium for
relativistic or nonrelativistic circular Coulomb orbits with a phase space
distribution $P(J_{2})=const\times\exp\left[  -J_{2}/\left(  \hbar/2\right)
\right]  $ which is independent of the orbital frequency. \ 3) Through lowest
nonvanishing order in the orbital eccentricity $\epsilon,$ radiation balance
holds for a charged particle in a Coulomb potential in classical zero-point
radiation, for both relativistic and nonrelativistic particle motions. \ 4) In
order to avoid self-ionization in elliptical orbits of high eccentricity,
relativity may be important for numerical simulations of the classical
hydrogen atom in classical zero-point radiation. \ The message here is very
different from the definitive statements regarding the failure of classical
electrodynamics which appear in the current modern physics textbooks. \

\end{document}